\documentclass[aps,prb,superscriptaddress,11pt]{revtex4-2}

\usepackage{amsmath}
\usepackage{amsfonts}
\usepackage{amssymb}
\usepackage{makeidx}
\usepackage{graphicx}
\usepackage{hyperref}
\usepackage[capitalize]{cleveref} 
\usepackage{bm} 
\usepackage{physics} 
\usepackage[linesnumbered,ruled,vlined]{algorithm2e}
\usepackage{qcircuit} 

\usepackage{setspace}
\usepackage{color}

\begin{document}

\title{Benchmarking Quantum Heuristics: \\Non-Variational QWOA for Weighted Maxcut}

\author{Tavis Bennett}
\email{tavis.bennett@research.uwa.edu.au}
\affiliation{Centre for Quantum Information, Simulation and Algorithms, School of Physics, Mathematics and Computing, The University of Western Australia, Perth, Australia}

\author{Aidan Smith}
\affiliation{Centre for Quantum Information, Simulation and Algorithms, School of Physics, Mathematics and Computing, The University of Western Australia, Perth, Australia}

\author{Edric Matwiejew}
\affiliation{Pawsey Supercomputing Research Centre, Perth, Australia}

\author{Jingbo Wang}
\email{jingbo.wang@uwa.edu.au}
\affiliation{Centre for Quantum Information, Simulation and Algorithms, School of Physics, Mathematics and Computing, The University of Western Australia, Perth, Australia}

\begin{abstract}
We present benchmarking results for the non-variational Quantum Walk Optimisation Algorithm (non-variational QWOA) applied to the weighted maxcut problem, using classical simulations for problem sizes up to $n = 31$. The amplified quantum state, prepared using a quadratic number of alternating unitaries, achieves a constant average-case measurement probability for globally optimal solutions across these problem sizes. This behaviour contrasts with that of classical heuristics, which, for NP-hard optimisation problems, typically exhibit solve probabilities that decay as problem size increases. Performance comparisons with two local-search heuristics on the same benchmark instances suggest that the non-variational QWOA may offer a meaningful advantage by scaling more favourably with problem size. These results provide supporting evidence for the potential of this quantum heuristic to achieve quantum advantage, though further work is needed to assess whether the observed performance scaling persists at larger problem sizes, and to confirm whether similar performance trends are observed for the other problem classes to which the non-variational QWOA is designed to generalise.
\end{abstract}

\maketitle

\vspace{-1cm}
\section{Introduction}
Many important real-world problems reduce to combinatorial optimisation, where the goal is to identify the best configuration from an exponentially large set of discrete alternatives. Examples include designing delivery routes, allocating limited resources, scheduling vehicles, and rostering personnel. Each candidate solution represents a combination of discrete decisions, and its quality is typically quantified by an objective function encoding cost, profit, or another real-world quantity to be optimised.

Among the various proposed applications of quantum computing, combinatorial optimisation stands out as both commercially valuable and potentially accessible to near-term hardware. Quantum algorithms are actively being developed in the hope of outperforming classical heuristics on hard optimisation problems, where even advanced classical methods return suboptimal solutions as problem size increases~\cite{abbas2024}. A foundational quantum algorithm in this space is the Quantum Approximate Optimisation Algorithm (QAOA) \cite{farhi2014quantum}, which prepares approximate solution states by using an alternating sequence of parameterised unitaries. This framework has been extended in several ways, including the Quantum Walk Optimisation Algorithm (QWOA) \cite{Marsh2019, hadfield2019quantum, marsh2020combinatorial, slate2021quantum, bennett2021quantum}, which replaces the standard QAOA mixing unitary with a continuous-time quantum walk (CTQW)~\cite{QWbook2014} on a circulant graph connecting solutions.

In this work, we study the performance of a recently introduced variant of QWOA known as the non-variational QWOA \cite{bennett2024non, bennett2024analysis}. Unlike typical variational quantum algorithms, which have an increasing number of parameters as problem size grows and suffer from issues such as barren plateaus and critical dependence on effective parameter initialisation \cite{bittel2021training,larocca2024review}, the non-variational QWOA prepares an amplified quantum state using only three parameters, regardless of problem size or circuit depth. In addition, it replaces the original QWOA's circulant graphs with problem-specific mixing graphs that capture solution structure. This enables the quantum walk to exploit structure and assist in significantly amplifying globally optimal solutions. For an in-depth analysis of the non-variational QWOA, including its design principles, theoretical underpinnings, and performance benchmarks, we refer the reader to \cite{bennett2024non, bennett2024analysis}. The analysis in \cite{bennett2024non, bennett2024analysis} highlights how the use of three parameters and problem-structured mixing unitaries enables efficient quantum state amplification, offering significant advantages in scalability and robustness over existing QAOA-type variational approaches.

We investigate whether the non-variational QWOA offers a performance advantage over classical heuristics for combinatorial optimisation. To do so, we benchmark its performance on randomly generated instances of the weighted maxcut problem: a problem where the goal is to partition the vertices of a weighted graph into two subsets such that the total weight of edges crossing the partition is maximised. Using simulations of problem sizes up to $n = 31$, we observe that the quantum heuristic maintains a constant average-case probability of measuring globally optimal solutions, while classical heuristics exhibit solve probabilities that decay with problem size. These findings provide preliminary evidence that the quantum heuristic may overcome limitations inherent to classical approaches, and support the hypothesis that the non-variational QWOA could offer a route to significant quantum advantage. The remainder of the paper is organised as follows: \cref{sec:classical_heuristics} introduces the classical heuristics and performance metrics; \cref{sec:benchmarking_methodology} outlines the simulation methodology; \cref{sec:results} presents results and comparative analysis; and \cref{sec:conclusion} discusses broader implications and concludes.

\vspace{-0.6cm}
\section{Classical heuristics and performance metrics}
\label{sec:classical_heuristics}
\vspace{-0.5cm}
In this paper, we will be benchmarking against the performance of local-search heuristics. In its simplest form, local search involves the iterative improvement of a candidate solution. Improvements of a candidate solution are constrained to small perturbations in the structure of the solution. This is usually described with reference to a neighbourhood structure, in which solutions that differ by a single perturbation are defined as neighbours. By making small incremental changes in solution configuration, local-search heuristics operate on the principle that solutions are not independent from each other, but rather they are related to each other through the decisions from which they are composed. Local-search heuristics are fast, effective, widely generalisable, and commercially competitive (solvers such as Hexaly and OR-Tools both use local search as part of their suite of algorithms and heuristics). In addition, local-search-based heuristics are expected to be effective on the same class of problems as the non-variational QWOA: the mixing graph is defined by a similar neighbourhood structure, and both approaches rely on structural correlations between neighbouring solutions and objective values.

\begin{figure}[htbp]
    \centering
    \includegraphics[width=0.49\columnwidth]{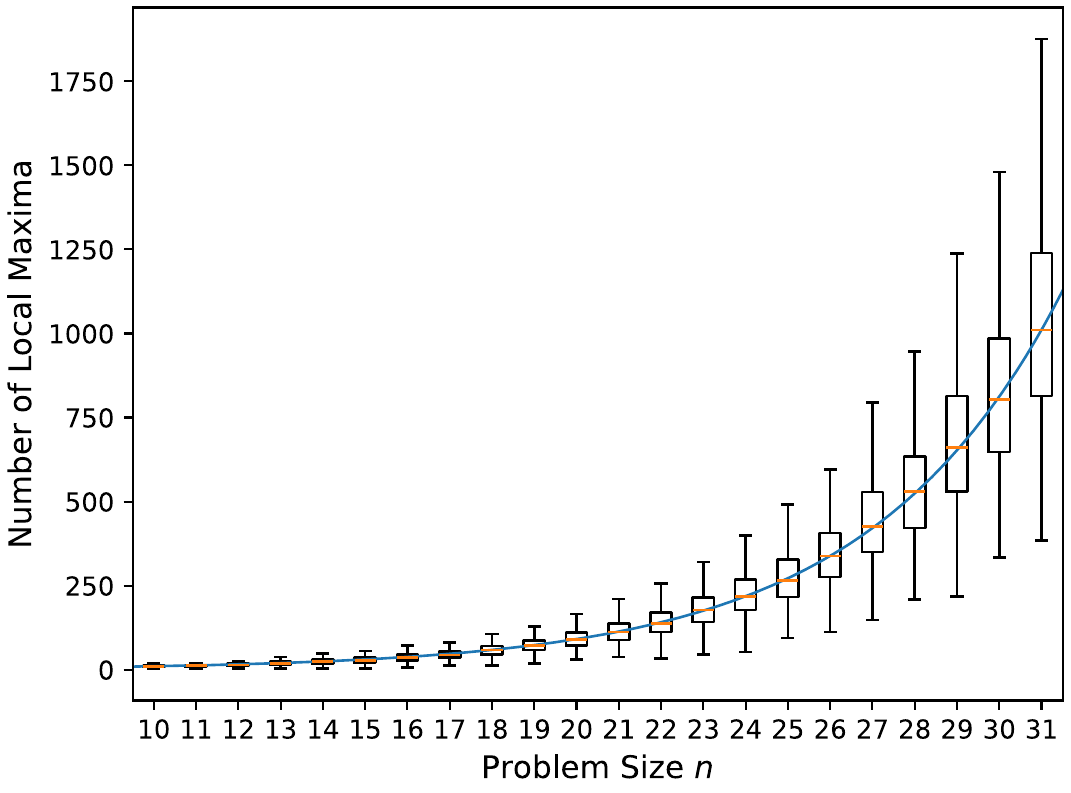}
    \includegraphics[width=0.48\columnwidth]{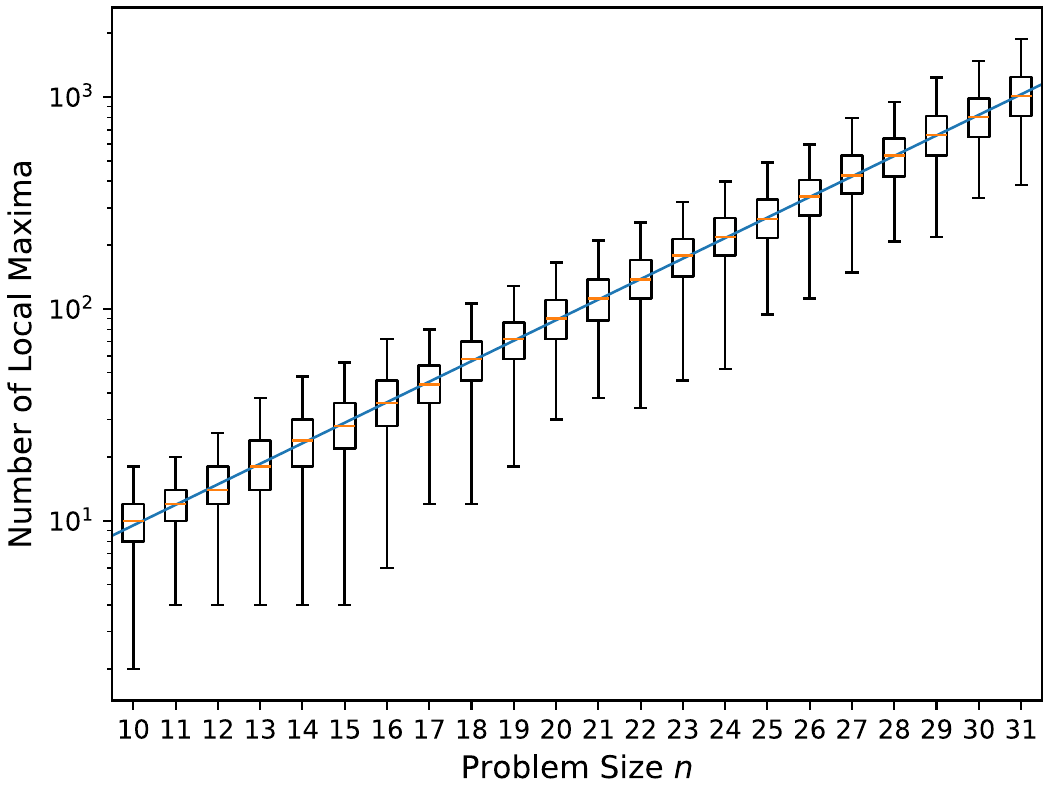}
    \caption[Counting Local Optima for weighted maxcut]{Distributions of the number of local optima for each problem size of the weighted maxcut benchmarking instances shown with (left) a linear scale, and (right) a log scale. The blue curve/line shows an exponential fit to the median values.}
    \label{fig:local_maxima_count}
\end{figure}

The primary weakness associated with local-search heuristics is that, by definition, they terminate at a solution that is only locally optimal, in a solution space where many such local optima are distributed throughout. In fact, the number of unique local optima present for a typical NP-hard optimisation problem grows exponentially with the size of the problem instance. To demonstrate this, we compute the number of local optima for each of the problem instances in the benchmark library described in \cref{sec:benchmarking_methodology}, and show the results as a function of problem size in \cref{fig:local_maxima_count} (for $10\leq n \leq 21$). The results clearly indicate a median value which grows exponentially in $n$. 

There are some ways around this---for example, iterated local search \cite{lourencco2003iterated}, similar to basin hopping in a continuous optimisation problem, makes use of perturbations that are larger than those which define the local neighbourhood, to escape from a local optimum, prior to restarting the local search process. Simulated annealing \cite{kirkpatrick1983optimization} allows for perturbations that decrease solution quality, so as to provide a mechanism to escape less favourable local optima. Tabu search \cite{glover1998tabu} performs repeated local search while avoiding previously explored regions, with the hope of finding improved local optima. 

There is also another important class of heuristics: those which are population-based \cite{beheshti2013review}, such as particle swarm optimisation, ant colony optimisation and genetic algorithms. These methods usually operate on a population of solutions, in contrast to the single ever-changing candidate solution used by a local-search-based heuristic, and as such, they tend to be more resilient to the local optima trap. These population-based heuristics are typically more computationally demanding than local-search approaches, often being implemented via parallel computation. 
While potentially quite powerful, both the local-search-based and population-based heuristics, regardless of their sophistication, ultimately face the same limitation; for a sufficiently large problem, the solution space becomes intractably large, with an exponential number of locally optimal regions beyond the reach of any polynomial-time heuristic.

There is a key feature of a quantum heuristic which makes it intrinsically different to a classical heuristic: the use of superposition allows for a global rather than a local approach, enabling the quantum heuristic to occupy and assess the entire solution space at once. In addition, the non-variational QWOA produces an interference effect which is largest for globally optimal solutions \cite{bennett2024analysis}. This means that even in the case of an exponentially large solution space, the amplification process is intrinsically biased towards globally optimal solutions. For this reason, one may expect the non-variational QWOA to be better able to produce optimal solutions than classical heuristics.

When assessing the performance of a heuristic, which by definition does not come with a performance guarantee, it is helpful to quantify average-case or typical performance. The probability (in the average case) of finding a globally optimal solution is expected to decay with growing problem size for a classical heuristic. As such, producing an average-case measurement probability of the globally optimal solution, which does not shrink with problem size, would constitute a significant performance advantage over classical heuristics. Note that classical heuristics do not exceed polynomial run-time, so of course, constant average-case measurement probability would have to be achieved with a polynomial runtime for a fair comparison. The purpose of this study is to assess, via large-scale simulations, whether this behaviour is observed for the non-variational QWOA. Specifically, we aim to verify whether a constant measurement probability for the optimal solution can be achieved using a number of iterations that scales polynomially with problem size.

\vspace{-0.6cm}
\section{Methods}
\label{sec:benchmarking_methodology}
\vspace{-0.5cm}
This study is restricted to binary problems and moderate problem sizes, due to the computational cost of simulating quantum systems. While the solution space for binary problems still grows exponentially with problem size, the base of this exponential is smallest for binary problems, allowing simulations over a broader range of sizes. Benchmarking non-binary problems is beyond the scope of this work, but will be an important focus of future research. Such studies will help assess whether the observed performance generalises to other problem structures where the non-variational QWOA relies on other custom mixing graphs.

In order to benchmark average-case performance, we focus on the weighted maxcut problem and generate a library of random problem instances. For each of the problem sizes $10 \leq n \leq 31$ we generate 1{\small,}000 weighted graphs using the Erdős–Rényi–Gilbert model. The selected range of problem sizes reflects the limitations of double-precision state-vector simulation imposed by the VRAM available to a single compute die of an AMD MI250X GPU~\cite{setonix2023}.

To solve a weighted maxcut instance to global optimality using the non-variational QWOA, we must reliably measure the optimal solution from an amplified state. We refer to the probability of measuring the optimal solution simply as the measurement probability, and we target an average-case measurement probability of $10\%$. Achieving a measurement probability on the order of $10\%$ essentially guarantees successful measurement of the optimal solution in the context of quantum algorithms in which circuits are typically measured thousands of times. The central purpose of the large-scale simulations is to determine how many iterations $p$ are required to meet this target, across a range of increasing problem sizes.

For a given problem instance and iteration count $p$, the measurement probability is taken to be that of the state which maximises the expected objective value. As detailed in \cref{fig:simulation}, the non-variational QWOA generates the layer-wise phase-separator angles $\boldsymbol{\gamma}=(\gamma_{1},\dots,\gamma_{p})$ and mixer durations $\mathbf{t}=(t_{1},\dots,t_{p})$ from only three continuous parameters $(\gamma,t,\beta)$, irrespective of $p$.  We use the limited-memory Broyden–Fletcher–Goldfarb–Shanno (BFGS) algorithm with box constraints---implemented via the Python package \textit{SciPy}---to tune these parameters and maximise the expectation value for each problem instance and iteration count.

\newcommand{\mycomment}[1]{\hfill $\triangleright$~\textit{#1}\\}
\begin{algorithm}[htbp]
    \caption{Simulated parameter optimisation for the non-variational QWOA}
    \KwIn{Weighted graph $G$ on $n$ vertices, iteration count $p$, initial guess $\mathbf{x}_0$}
    \KwIn{Parameter bounds: $\gamma \in [0,5]$, $t \in [0,0.7]$, $\beta \in [0,0.5]$}
    \KwIn{Circuit parameters: $\{(\gamma_k, t_k)\}_{k=1}^p$}
    \SetKwFunction{FObjective}{Objective}
    \SetKwFunction{FBuild}{BuildQWOACircuit}
    \SetKwFunction{FSim}{StatevectorSim}
    \SetKwFunction{FMinimize}{minimize}
    \SetKwFunction{FObjective}{Objective}
    \SetKwFunction{FComputeCuts}{f}
    \SetKwFunction{Fstd}{std}
    \BlankLine
    $\mathbf{C} \gets \FComputeCuts(G)$ \mycomment{Compute maxcut objective values}
    $\sigma \gets \Fstd(\mathbf{C})$ \mycomment{Compute standard deviation}
    $q_c \gets \FBuild(n, p, \{\gamma_k, t_k\})$ \mycomment{Build the parameterised circuit}
    \BlankLine
    \SetKwProg{Fn}{Function}{:}{}
    \Fn{\FObjective{$\mathbf{x}=(\gamma, t, \beta)$}}{
        \For{$k = 1$ \KwTo $p$}{
            $\gamma_k \gets \frac{\gamma}{\sigma} \left( \beta + \frac{k-1}{p-1}(1 - \beta) \right)$ \\
            $t_k \gets t \left( 1 + \frac{k-1}{p-1}(\beta - 1) \right)$}
        $q_c^\ast \gets q_c[\forall j: \gamma_j \leftarrow \gamma_j,\ t_j \leftarrow t_j]$ \mycomment{Bind parameters}
        $\bm{\psi} \gets \FSim(q_c^\ast)$ \mycomment{Simulation via Qiskit Aer simulator}
        $\mathbf{P} \gets |\bm{\psi}|^2$ \\
        \Return $-\sum_b P_b C_b$ \mycomment{Negative expectation value to be minimised}}
    \BlankLine
    $(\gamma^\ast, t^\ast, \beta^\ast) \gets \FMinimize(\FObjective, \mathbf{x}_0, \text{bounds})$ \mycomment{L-BFGS optimiser}
    \label{fig:simulation}
\end{algorithm}

\vspace{-0.5cm}
Starting with $n = 10$ and $p = 2$, simulations are repeated to tune the parameters and produce an amplified state for each of the 1{\small,}000 problem instances, enabling the final measurement probabilities to be extracted for each problem instance. We increment the number of iterations until the mean measurement probability across problem instances exceeds the target value. This process is repeated for larger problem sizes, adaptively selecting iteration counts to focus only on those $p$ values near the target threshold, thereby minimising simulation overhead.

For a size $n$ unconstrained binary optimisation problem, such as maxcut, the non-variational QWOA is executed with an $n$-qubit quantum circuit, as solutions correspond one-to-one with computational basis states. Consider a problem instance defined by a weighted graph with weights $w_{ij}$ assigned to each pair of connected vertices $i$ and $j$. The quantum circuit necessary to implement the non-variational QWOA for this (and any QUBO) problem is the same as that used for the QAOA, with one qubit assigned to each vertex and initialised with Hadamard gates in the equal superposition state.

\begin{figure}[htbp]
    \centering
    \[ \Qcircuit @C=2em @R=0.7em {
    \lstick{} & \qw & \gate{R_x(2t)} & \qw \\
    \lstick{} & \qw & \gate{R_x(2t)} & \qw \\
    \lstick{} & &  \vdots & \\
    \lstick{} & & & \\
    \lstick{} & \qw & \gate{R_x(2t)} & \qw }%
    \hspace{4cm}\Qcircuit @C=2em @R=3em {
    \lstick{\text{qubit } i} & \ctrl{1} & \qw & \ctrl{1} & \qw \\
    \lstick{\text{qubit } j} & \targ &  \gate{P(-\gamma w_{ij})} & \targ & \qw \\
    & & & & } \]
    \vspace{-0.9cm}
    \caption[Quantum circuits for simulation of maxcut and other QUBOs]{Quantum circuits for (left) the mixing unitary with mixing time $t$ and (right) the phase separation specifically associated with the edge between vertex $i$ and $j$ with edge weight $w_{ij}$ and phase-separation strength $\gamma$.}
    \label{fig:quantum_circuits}
\end{figure}
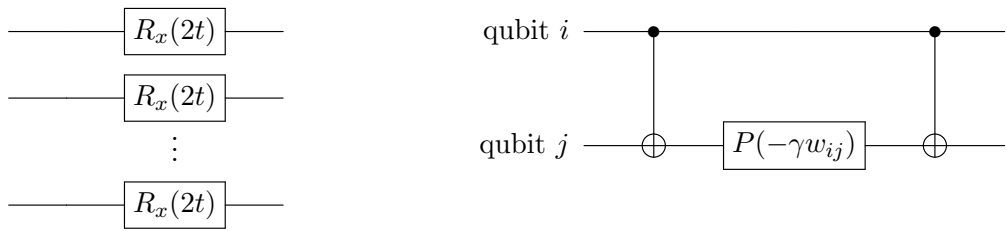

The mixing unitary is performed by single qubit rotations as shown in \cref{fig:quantum_circuits}~(left), and the phase separation unitary is composed of two-qubit operations, one for each pair of connected vertices, as shown in \cref{fig:quantum_circuits}~(right). This interaction can equivalently be implemented using an RZZ gate.  Simulating the resulting quantum circuit is achieved using the Qiskit Aer simulator with a GPU backend and without a noise model. 

\vspace{-0.4cm}

\vspace{-1.2cm}
\section{Results and analysis}
\label{sec:results}
\begin{figure}[htbp]
    \centering
    \includegraphics[width=0.96\linewidth]{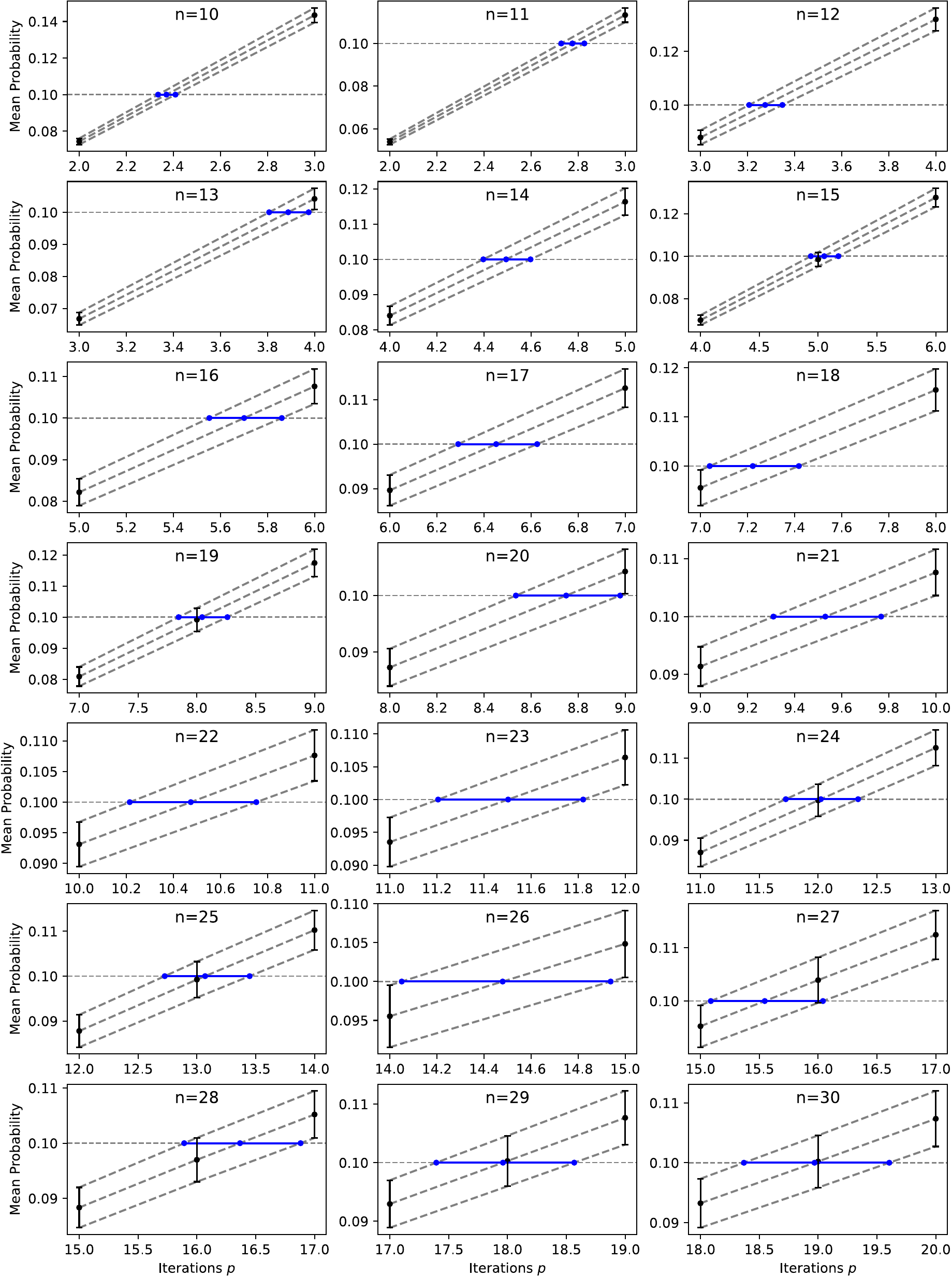}
    \caption[Benchmarking: Simulation results]{Simulation results, indicating typical performance for weighted maxcut problem instances of size $10 \leq n \leq 30$. The black error bars indicate mean measurement probability with 95\% confidence intervals, and the blue points indicate the interpolated number of iterations required for target 10\% measurement probability.}
    \label{fig:required_iterations_interpolation}
\end{figure}

\vspace{-0.3cm}
The simulation results for all problem instances from $n=10$ to $n=30$, inclusive, are shown in \cref{fig:required_iterations_interpolation}. The mean measurement probability is shown with black error bars indicating 95\% confidence intervals, computed from simulation results for the set of 1000 randomly generated problem instances at the associated iteration count $p$. Due to the discrete nature of the iteration count $p$, we are interested in interpolating and estimating the exact non-integer number of iterations required to exactly meet the target measurement probability. This is achieved via linear interpolation, and the confidence region of this estimate is taken from the linear interpolation of the 95\% confidence intervals, shown in blue.

\begin{figure}[htbp]
    \centering
    \includegraphics[width=0.6\linewidth]{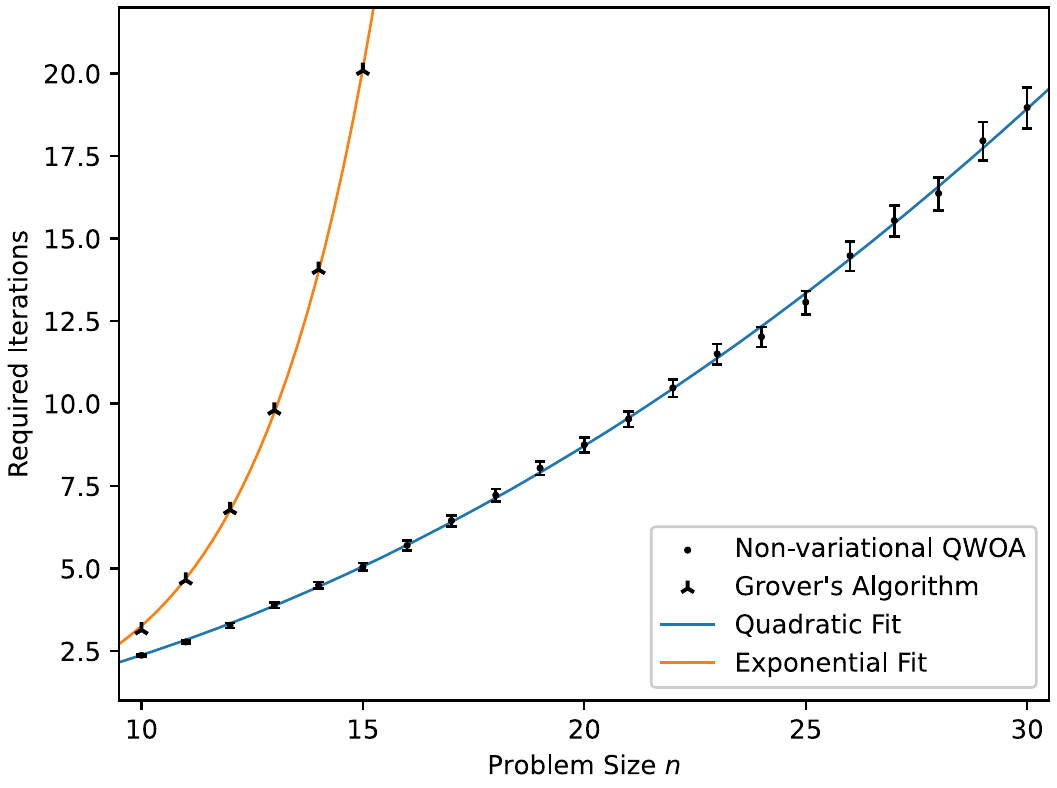}
    \caption[Benchmarking: Quadratic iteration count]{Estimated number of iterations $p$ required to achieve target 10\% measurement probability for both the non-variational QWOA and Grover's algorithm, along with a quadratic and exponential fit, respectively.}
    \label{fig:required_iterations}
\end{figure}

\newpage The estimated number of iterations required to produce a mean measurement probability equal to the target 10\%, shown in blue in \cref{fig:required_iterations_interpolation}, are compiled in a single plot in \cref{fig:required_iterations}. Importantly, a quadratic fit,
\begin{equation}
    \label{eq:required_iterations}
    p(n) = 0.019 n^2 + 0.053 n - 0.092,
\end{equation}
is shown in blue which agrees very closely with the estimated values. As a comparison, the iterations of Grover's algorithm required to achieve the same target measurement probability are also shown, with an associated exponential fit. This indicates that the non-variational QWOA is able to exploit problem structure to provide significant speedup relative to the quantum unstructured search. However, most importantly, the number of iterations required to produce constant average-case measurement probability scales quadratically in the problem size, at least over the range of simulated problem sizes. 

\begin{figure}[htbp]
    \centering
    \includegraphics[width=0.65\linewidth]{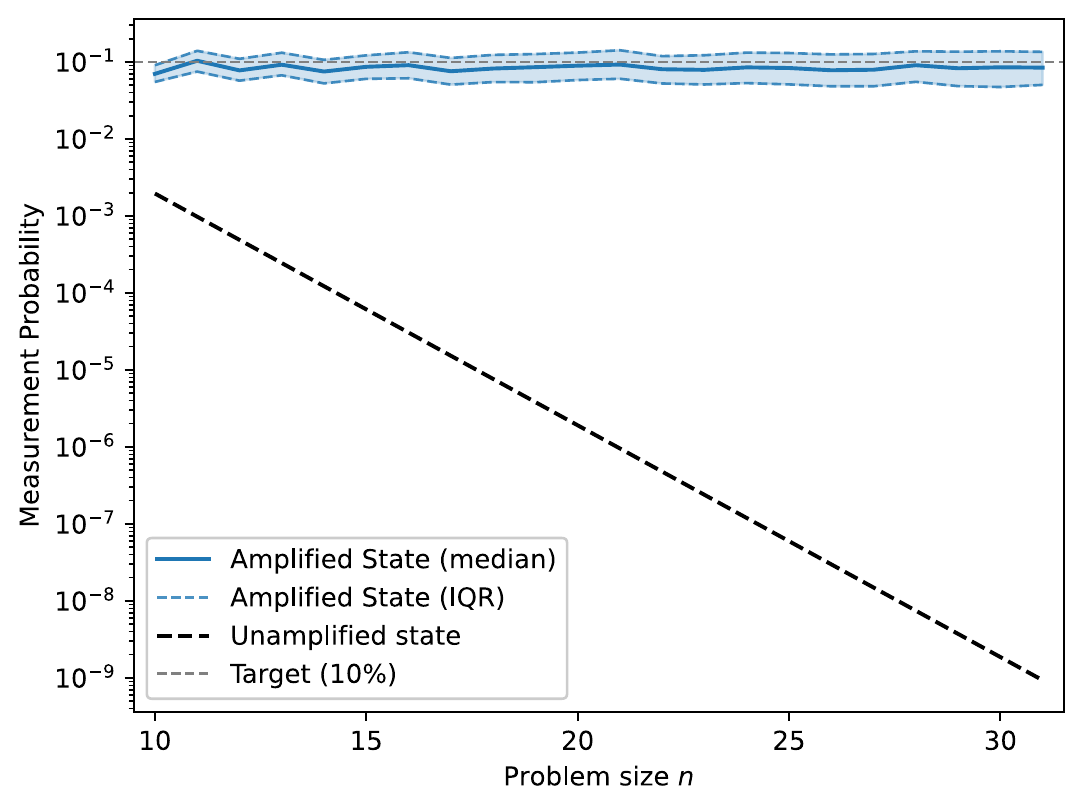}
    \caption{Typical performance of the non-variational QWOA applied to weighted maxcut on random graphs. Optimal solutions are amplified to measurement probabilities around $10\%$ in the typical case, as indicated by the median and interquartile range. The unamplified measurement probability is indicated with the dashed black line.}
    \label{fig:Amplification_to_the_measurable_regime}
\end{figure}

The typical performance of the non-variational QWOA, for $10 \leq n \leq 31$, is shown in \cref{fig:Amplification_to_the_measurable_regime}, where iteration counts are given by \cref{eq:required_iterations}, rounded to the nearest integer. These results demonstrate that the heuristic consistently amplifies globally optimal solutions up to measurement probabilities around the target $10\%$ using only quadratic circuit depth. This finding is significant: if the observed behaviour extrapolates to larger problem sizes, it suggests that the quantum heuristic may be capable of solving the weighted maxcut problem on random graphs---an NP-hard optimisation problem---to global optimality, in the average case. Such performance is not achievable with classical heuristics under polynomial time constraints. While extrapolation beyond the simulated range introduces uncertainty, the interference effect at the core of the non-variational QWOA, as described in \cite{bennett2024analysis}, relies on statistical assumptions that are increasingly valid for larger problem sizes. This supports the plausibility of the observed scaling persisting for larger problem sizes.

Note that for the size 31 problems, a typical measurement probability of around 10\% corresponds to an amplification of the optimal solution's measurement probability by over 100 million relative to the equal superposition state. This compares to an amplification of 1{\small,}681 (5 orders of magnitude difference) achieved by Grover's algorithm with the same number of iterations ($p=20$). In addition, the constant measurement probability indicates an amount of amplification which grows exponentially in problem size, despite requiring only a quadratic circuit depth. The non-variational QWOA is clearly able to exploit problem structure to achieve advantage over a quantum unstructured search. It also has a remarkably low time-complexity, which we can show by making a comparison with local-search heuristics.

\begin{figure}[htbp]
    \centering
    \includegraphics[width=0.49\columnwidth]{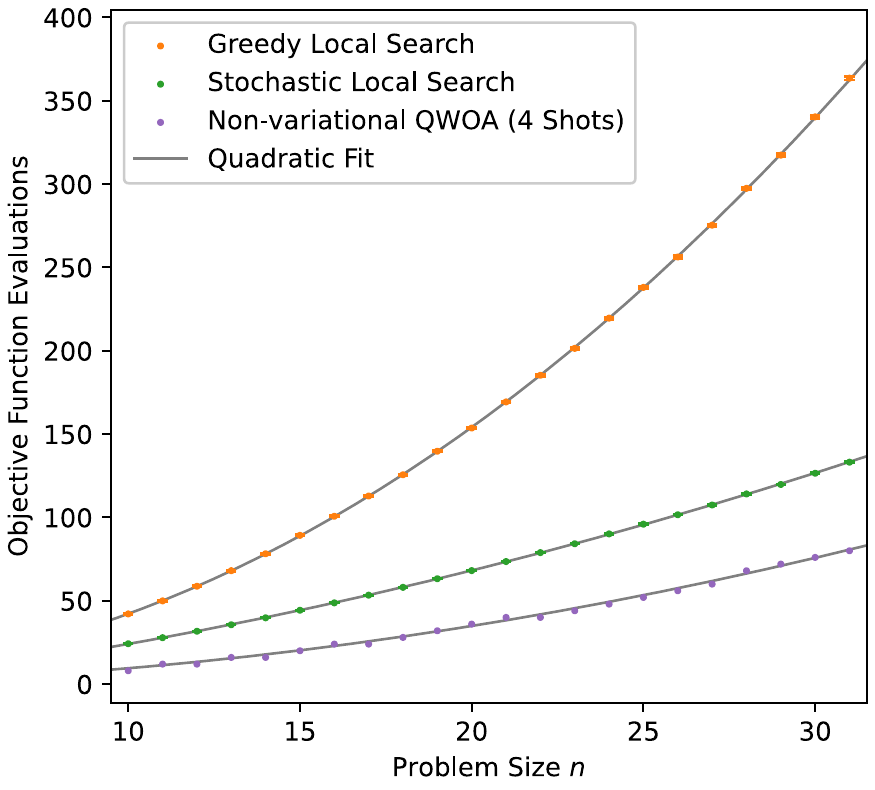}%
    \includegraphics[width=0.49\columnwidth]{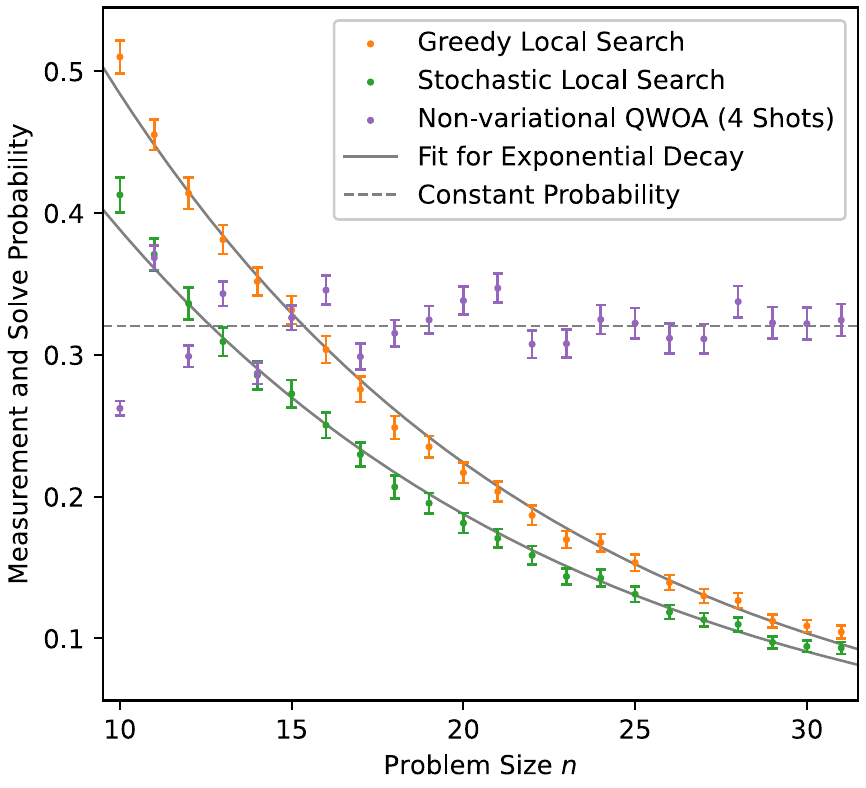}
    \caption[Benchmarking: Comparison between local search and the non-variational QWOA]{A performance comparison between the local-search heuristics and the non-variational QWOA. (left) The mean number of iterations or objective function evaluations required for each heuristic, which is proportional to the computational time required. (right) Mean measurement and solve probabilities, indicating constant probability for the quantum heuristic vs an approximately exponentially decaying probability for the local-search heuristics, as anticipated. All error bars show $95\%$ confidence intervals.}
    \label{fig:comparison_with_LS}
\end{figure}

\cref{fig:comparison_with_LS} directly compares the performance of the classical local-search heuristics with that of the non-variational QWOA. For the quantum heuristic, the number of iterations is taken from \cref{eq:required_iterations}, rounded to the nearest integer. Measurement probabilities fluctuate above and below the $10\%$ target, depending on whether the iteration count was rounded up or down. Each result shown is based on four repeated measurements of the amplified quantum state (i.e. four shots).

To evaluate and compare performance, we report three key metrics: the measurement probability, defined as the average-case probability of sampling a globally optimal solution from four measurements of the amplified quantum state; the solve probability, defined as the average-case fraction of classical heuristic runs that return a globally optimal solution; and the number of objective function evaluations, which provides an indication of time-complexity. For the classical heuristics, the number of function evaluations is effectively proportional to the time required for termination of the local search. For the quantum heuristic, each iteration involves a phase-separation unitary that encodes the objective function, with equivalent time complexity to a classical objective function evaluation. Hence, $4p$ effectively quantifies the number of function evaluations required to prepare and measure the amplified state four times.

\cref{fig:comparison_with_LS}(left) shows that each of the classical heuristics we used exhibit quadratic growth in objective function evaluations with problem size, but the quantum heuristic requires the fewest. \cref{fig:comparison_with_LS}(right) compares the measurement and solve probabilities. As expected, the solve probabilities for the local-search heuristics decay approximately exponentially, while the quantum heuristic appears to maintain a constant measurement probability, even as problem size increases.

\section{Discussion and outlook}
\label{sec:conclusion}
While this study benchmarks the non-variational QWOA against basic local-search heuristics, one might reasonably ask whether stronger classical alternatives could have been selected. For instance, breakout local search \cite{benlic2013breakout} (a variant of iterated local search) likely achieves perfect solve probability at these simulable problem scales, particularly with sufficient breakouts/perturbations. However, such a heuristic would come with higher runtime and would remain fundamentally constrained by local access to the solution space. Although its perturbation strategy generally helps escape local optima, it does not eliminate the underlying problem: the presence of exponentially many locally optimal regions in the solution space. At larger problem sizes, such methods are still expected to fail with high probability. In this context, comparing against basic local search is appropriate and deliberate, as it clearly illustrates the distinction between classical heuristics limited to local navigation and the quantum heuristic’s global, interference-driven mechanism.

Indeed, the results presented here suggest that the non-variational QWOA does not merely sidestep the local-optima trap but overcomes it through fundamentally non-classical means. By preparing a superposition over the entire solution space and employing a global interference process, the heuristic achieves preferential amplification of globally optimal solutions. This behaviour appears robust to increasing problem size---at least over the simulable regime---and points to a genuine performance advantage over classical methods constrained to local exploration.

At sufficient problem size, classical heuristics constrained to polynomial runtime exhibit a vanishing solve probability. If a quantum heuristic were capable of sustaining a constant average-case measurement probability for arbitrary problem size (subject to polynomial runtime), this would constitute a quantum advantage. Thus, observing this behaviour over the simulable regime, as we have demonstrated here, is a necessary condition for quantum advantage. However, it is not a sufficient one. Some readers may rightly hesitate to interpret these results as conclusive evidence of quantum advantage, given the limited scale of simulations. A sufficient condition would require demonstrating the same performance for arbitrarily large problem sizes---something which is infeasible via simulation and well beyond the reach of near-term quantum hardware. Until such implementations are feasible, the case for quantum advantage must instead rest on the accumulation of supporting evidence, such as that presented here.

Finally, while this paper focuses on unconstrained binary optimisation, the non-variational QWOA applies to a much broader class of problems. The same underlying interference process supports its application to constrained and non-binary optimisation, with generalisation made possible through custom penalty functions and mixing graphs. It is anticipated that the observed performance trends will generalise, but confirming this remains an important direction for future work. Together, these findings motivate continued investigation of quantum heuristics for real-world combinatorial optimisation problems.

\section*{Acknowledgements}

This research was undertaken with the assistance of resources and services from the National Computational Infrastructure (NCI), supported by the Australian Government, and resources provided by the Pawsey Supercomputing Research Centre’s Setonix Supercomputer, funded by the Australian Government and the Government of Western Australia. TB and AS acknowledge support from Australian Government Research Training Program Scholarships. TB further acknowledges the Jean Rogerson Postgraduate Scholarship, and AS the Bruce and Betty Green Postgraduate Research Scholarship, all at The University of Western Australia.

\bibliography{refs}

\end{document}